\documentstyle[12pt,epsf]{article}

\newcommand{\xbj}{x_{B}}
\newcommand{\dt}{\Delta_T}

\newcommand{\M}[4]{M^{#1, #2}_{#3, #4}}     
\newcommand{\MM}[2]{\tilde{M}^{#1, #2}}     
\newcommand{\tvc}{{\cal T}_{\it VCS}}       
\newcommand{\tbh}{{\cal T}_{\it BH}}        

\newcommand{\bcn}{\begin{center}}
\newcommand{\beq}{\begin{equation}}
\newcommand{\beqn}{\begin{eqnarray}}
\newcommand{\ecn}{\end{center}}
\newcommand{\eeq}{\end{equation}}
\newcommand{\eeqn}{\end{eqnarray}}

\newcommand{\lsim}{\raisebox{-.5ex}{\footnotesize$
     \,\:\stackrel{\textstyle<}{\sim}\,\:$}}

\newcommand{\eqpt}{\hspace{6pt}.\hspace{6pt}}  
\newcommand{\eqcm}{\hspace{6pt},\hspace{6pt}}
\newcommand{\GeV}{\mbox{\ GeV}}
\newcommand{\MeV}{\mbox{\ MeV}}
\newcommand{\gstar}{\gamma^\ast}
\newcommand{\re}{\mbox{Re}}                 
\newcommand{\im}{\mbox{Im}}
\newcommand{\weight}[1]{\langle\langle \, #1 \, \rangle\rangle}

\begin{document}

\begin{flushright}
  CPT--S511--0697 \\
  NIKHEF--97--024
\end{flushright}
\vspace{2\baselineskip}

\begin{center}
{\Large \bf Testing the handbag contribution to exclusive virtual Compton
scattering} \\
\vspace{3\baselineskip}
{\large Markus Diehl$^{{\protect 1}}$, Thierry Gousset$^{{\protect
2}}$, Bernard  Pire$^{{\protect 1}}$and John P. Ralston$^{{\protect
3}}$} \\
\vspace{2\baselineskip}
{\it 1. CPhT{\hspace{3pt}\footnote {Unit\'e propre 14 du Centre
National de la Recherche Scientifique.}}, Ecole Polytechnique, 91128
Palaiseau, France. \\
2. NIKHEF, P. O. Box 41882, 1009 DB Amsterdam, The Netherlands. \\
3. Dept.\ of Physics and Astronomy, University of Kansas, Lawrence,
KS~66045, USA.}
\\
\vspace{8\baselineskip}
\large {\bf ABSTRACT}
\end{center}

\begin{center}
  \parbox{0.9\textwidth}{We discuss the handbag approximation to
    exclusive deep virtual Compton scattering. After defining the
    kinematical region where this approximation can be valid, we
    propose tests for its relevance in planned electroproduction
    experiments, $e + p \to e + p + \gamma$. We focus on scaling laws
    in the cross section, and the distribution in the angle between
    the lepton and hadron planes, which contains valuable information
    on the angular momentum structure of the Compton process. We
    advocate to measure weighted cross sections, which make use of the
    data in the full range of this angle and do not require very high
    event statistics.}
\end{center}

\newpage

\noindent
1. Exclusive virtual Compton scattering (VCS) has some very
interesting features. The subject has a long history, being examined
early in the development of the parton model~\cite{oldVCS}.
Recently~\cite{Ji,Rad,CHR} there has been considerable renewed
interest with the realization that off-diagonal correlations of quark
operators in proton states might be measurable with VCS. They
generalize the parton distributions of deep inelastic scattering (DIS)
but, being off-diagonal matrix elements, do not have a probability
interpretation. In fact, these ``off-diagonal parton distributions''
have a long history~\cite{OFPD}. Following the work of Ji, Tang and
Hoodbhoy~\cite{HoodJi}, Ji pointed out the relevance of VCS to getting
information about quark orbital angular momentum~\cite{Ji}. Some
analysis in the operator product expansion has been done in
~\cite{Wat,Chen}.

\begin{figure}[ht]
  $$\epsfysize 4cm\epsfbox{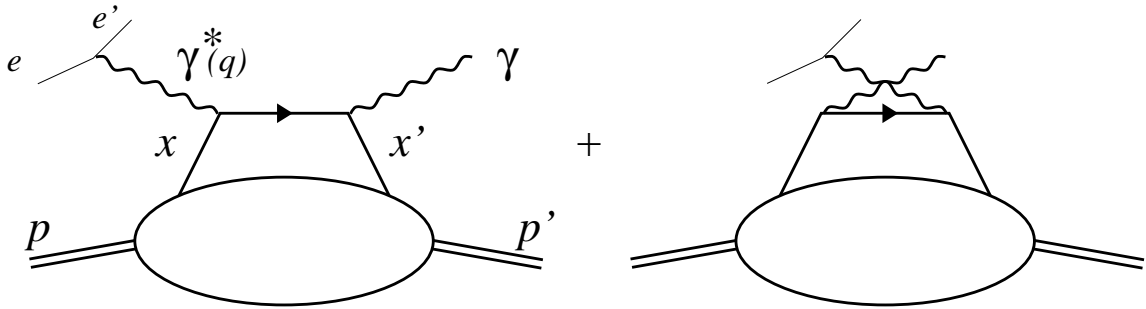}$$
  \caption{\label{handbag}The handbag Born diagrams for virtual
    Compton scattering. Here $x$ and $x'$ are light cone fractions of
    the quark with respect to the incoming proton momentum.}
\end{figure}

In DIS the diagrams that contribute at leading twist to the cross
section factorize into a hard perturbative and a soft nonperturbative
part, which are connected by only two quark or two gluon legs, where
furthermore the corresponding loop integral only runs over the parton
momentum fraction $x$. This defines the handbag approximation. Such
diagrams evidently also contribute to the VCS amplitude. We show in
Fig.~\ref{handbag} the handbag diagrams at Born level. To higher order
in $\alpha_S$, one has quark diagrams with radiative corrections to 
the hard scattering parts, as well as diagrams where the two partons 
attached to the soft blob are gluons.

There is no doubt that the handbag diagrams are but a subset of those
contributing to the process. Considerable theoretical effort has
already gone into exploiting the properties of the handbag model, for
instance, studying logarithmic scaling violations~\cite{Ji,Rad,Chen}.
A much different question is to critically ask how well one can trust
to the dominance of the handbag model in the first place.  On the
practical side, the new elements of VCS, off-diagonal in momentum and
spin, with the complication of a real photon in the final state, are
not dynamically similar to inclusive electroproduction but rather to
other exclusive reactions. In particular the size of (higher twist)
corrections to the handbag contribution, such as for instance the pion
exchange process discussed in~\cite{Afa}, really has to be studied
anew empirically: the values of $Q^2$ where the handbag approximation
is satisfactory need not be the same in deep inelastic scattering and
in VCS. At this point in time, not enough is known to make iron-clad
statements directly from theory.  As we argue below, the scope of the
handbag is most probably limited to a restricted kinematic region.
But whatever theory statements are made, the function of experimental
science is to test the important ideas systematically. A very diverse
experimental program is now being discussed for JLAB, HERA, CERN and
the ELFE project~\cite{exp}.

The dynamics of VCS is quite rich, and very different physical
processes will occur in different kinematic regions. No single
mechanism can be expected to describe all regions. Our approach here
is to ask: how can one test experimentally the dominance of the
general model, allowing for lack of information about the new
parton distributions about which almost nothing is known so far? To
proceed we will first focus on the kinematic regions where the model
might work, and then find some tests based on general principles.

\vspace{\baselineskip}
\noindent
2. Let us then have a closer look at the kinematics of the process
\begin{equation}
  \label{gammastarp}
  \gstar(q) + p(p) \to \gamma(q') + p(p')  \eqcm
\end{equation}
with momenta given in parentheses. We use the conventional variables
\begin{equation}
Q^2 = -q^2 \eqcm W^2 = (p+q)^2 \eqcm \xbj = Q^2/ (2 p \cdot q) \eqcm
\end{equation}
and consider only $Q^2 \gg 1 \GeV^2$ so that pQCD may be applied.
Taking $Q^2$ large while keeping $\xbj$ fixed corresponds to the
``Bjorken limit'', but for our reaction is too broad a range for the
handbag's application. In fact, much of the physics hinges on the
different components of the four-vector $\Delta^\mu = (p - p')^\mu$.
Note that $\Delta^2 =t$ is the additional Mandelstam variable
necessary to fix the kinematics. We will always work in the $\gstar p$
center of mass, with the positive $z$-axis given by the virtual photon
momentum, and introduce $\dt$ as the transverse momentum transfer from
the initial proton to the final one. For the scattering angle $\theta$
between $p$ and $p'$ we have
\begin{equation}
  \label{sintheta}
  \sin\theta = \frac{2 \dt W}{W^2 - m^2}  \eqpt
\end{equation}
At $\theta = 0$ the invariant $t$ attains its kinematical limit, given
by
\begin{equation}
  \label{tmin}
  t_0 = - m^2 \frac{\xbj^2}{1 - \xbj + \xbj m^2 /Q^2}
\end{equation}
up to relative corrections of order $\xbj m^2 /Q^2$, where $m$ is the
proton mass.

Now let us turn to different kinematic regions at large $Q^2$ and the
relevance of the handbag model to each:
\begin{itemize}
\item Resonances: there is an ``exceptional'' region, where $W$ is in
  the range of resonance masses and $\xbj \approx 1$. We remark that
  kinematics then constrains $- t$ to be of order $Q^2$,
  cf.~(\ref{tmin}), while $\dt$ is restricted to be of order $m$.
  Physically, in this region a rather soft photon is emitted in a
  nearly elastic absorption of a large $Q^2$ photon by the proton.  It
  is not a good bet that in the resonance region the long-time process
  of photon emission from the system is going to be well described by
  the impulse approximation used in the diagrams of
  Fig.~\ref{handbag}. Remember that the impulse approximation occurs
  when the ``plus'' component $k^+$ of the parton momentum (in our
  frame, where $p^{-}$ is large) is integrated over, setting relative
  light cone times in the correlation functions to zero. We emphasize
  that this situation is qualitatively different from the case of DIS,
  where the quasi-elastic region $\xbj \to 1$ is already rather
  special~\cite{Roberts}.

  In the following we exclude this region from our study and require
  $W^2 \gg 1 \GeV^2$, which is tantamount to
  \begin{equation}
    \label{restrictx}
    1 - \xbj \gg m^2 / Q^2  \eqpt
  \end{equation}
\item The fixed angle region: in the case where $\dt$ is of order $Q$,
  all invariants $Q^2$, $W^2$, $- t$ and $- u = - (p - q')^2$ are
  large compared to $1 \GeV^2$. This is called the ``fixed angle
  limit'', cf.~(\ref{sintheta}), and is an example of a short distance
  process governed by one and only one large scale $Q^2$. The handbag
  cannot be dominant in this region, which receives many roughly equal
  contributions from numerous quark counting-type diagrams, where only
  the three-quark Fock state of the proton is taken into account and
  where $\gstar$ and $\gamma$ may couple to different quarks, in
  contrast to the handbag diagrams of Fig.~\ref{handbag}.  Extensive
  calculations within pQCD have been performed for this
  region~\cite{fixed,KSG}. The fixed angle
  limit is an example of kinematics ``in the Bjorken limit'' which
  must be excluded to narrow down where the handbag might be useful.
  
\item The forward but large $\dt$ region: for sufficiently large $Q$
  there is a transition region where $\dt$ is much smaller than $Q$
  but still large compared with $1 \GeV$. From~(\ref{sintheta}) we see
  that this corresponds to forward scattering, $\theta \approx 0$. Let
  us argue why the handbag approach cannot apply here. In the $\gstar
  p$ frame the model describes the scattering of one quark out of the
  initial fast proton into the final one. The total momentum transfer,
  and in particular its transverse component $\dt$ is carried by this
  one quark. The scattered quark must re-coalesce with the spectator
  partons to form the outgoing proton in soft, non-perturbative
  processes, which are parameterized by the lower blob, or in other
  words the parton distributions. Yet one cannot ask soft processes to
  bend the final state through a $\dt$ of several GeV.
  
  One must instead model those scatterings perturbatively, directly
  inserting the momentum transfers to the spectators as is done in the
  fixed angle region --- although the presence of two different hard
  scale $\dt$ and $Q$ will make the problem more complicated. We point
  out that this problem has some similarity with hadron-hadron
  scattering at large c.m.\ energy and $- t$ of several
  $\GeV^2$~\cite{Soti}.  The presence of two scales poses a question
  of Sudakov effects although processes with photons are known to be
  much less sensitive than hadron exclusive reactions~\cite{farster}.
\item The small $\dt$ region: for the handbag dominance to be used one
  needs a region of $Q^2$ and $W^2$ large, where in addition $\dt$ is
  small enough so that the unscattered spectator partons have a good
  overlap with the scattered proton, and ``flow right through''. In
  practice, a usual criterion for constituents that live inside a
  proton is $\dt \lsim 300 \MeV$, but one might stretch it to as much
  as $\dt \lsim 1 \GeV$.  We call this the ``small $\dt$ region'' to
  emphasize that ``forward region'' as specified by the scattering
  angle $\theta$ does not take into account that partons have finite
  transverse momentum inside hadrons.

  In this region we have
  \begin{equation}
    \label{deltat}
    \dt^2 = (1 - \xbj) (t_0 - t)
  \end{equation}
  up to relative corrections of order $\xbj m^2 /Q^2$, so that $t$ is
  close to $t_0$ given in~(\ref{tmin}).  Notice that while $\dt$ has
  to be small the longitudinal components of the momentum transfer
  $\Delta^\mu$ are not: its ``minus'' component is $\Delta^{-} = \xbj
  p^{-}$ in our kinematic limit and thus of order $Q$.
\end{itemize}

We have, then, delineated the domain where the interesting physics of
the handbag, and the chance to measure the off-diagonal parton
distributions mentioned in the introduction, can be approached. It
turns out that in this kinematics there are rather simple tests which
could tell us whether the handbag is a good approximation or not. To
formulate these tests we will now turn to what the handbag model
predicts for angular momentum selection rules of the $\gstar p \to
\gamma p$ process.

\vspace{\baselineskip}
\noindent
3. Consider the perturbative Compton scattering of a virtual photon on
a free quark target in the high energy limit. The general expression
is complicated, but in a frame where all particles are fast and in the
region of nearly collinear kinematics, with transverse momenta much
smaller than $Q$, a short calculation reveals a handsome
simplification. In the collinear limit the process has the remarkable
property of $s$-channel helicity conservation for the photon. Note
first that in collinear kinematics angular momentum conservation is
the same as spin conservation. Next, recall that a light quark's
helicity is conserved (this is perturbative chiral symmetry). The
quark cannot absorb a longitudinal photon and emit another transverse
one at all. As for absorbing a transverse photon, the quark
flowing through (without flipping its helicity) then cannot change the
photon's helicity, leading to the $s$-channel photon helicity
conservation.

This argument, in the spirit of the old-fashioned parton model,
summarizes the spin selection rules of the handbag diagrams in
Fig.~\ref{handbag}. Recall that in order to get the $\gstar p \to
\gamma p$ amplitude parameterized by the off-diagonal quark correlation
functions of~\cite{Ji} one uses the impulse approximation and
integrates over the $k^{+}$ and the $k_T$ of the partons in the
handbag loop. The leading term in the expansion of the result in
powers of $1 /Q$ amounts to scattering on a free, on-shell parton,
just as in DIS. The off-diagonal correlation functions 
depend on three kinematical variables, which can be chosen as $x$,
the light cone fraction of the quark taken out of the target, $x'$,
the fraction of the quark put back into the scattered proton (both
fractions with respect to the initial proton momentum), and $\dt$ or
$t$. Kinematics fixes the difference $x - x'$ to be $\xbj$ in our
process.

In the diagrams of Fig.~\ref{handbag} the integration over the light
cone fractions falls into three distinct regions: $x$ and $x'$ are
either both positive, or both negative, or $x$ is positive and $x'$
negative. In the first case our above argument directly gives
$s$-channel helicity conservation for the photon. In the second case
one can, as one does in DIS, re-interpret the ``backward moving'' quarks
as forward moving antiquarks and apply our argument
to Compton scattering on an antiquark. In the third case, which does
not have an analogue in DIS, one can again map the backward moving
$x'$-quark onto an antiquark. The partonic process then is the
collision of a quark-antiquark pair with a virtual photon, giving a
real photon in the final state. With quark helicity conservation and
collinear kinematics the $q \bar{q}$-pair has zero spin along the
collision axis, so that again the photon helicity is conserved.

We have then that in the handbag approximation of Fig.~\ref{handbag}
the photon does not change helicity to leading order in $1 /Q$. At
$\dt = 0$ the proton helicity can then not be flipped either because
of angular momentum conservation. For finite small $\dt$, however, the
handbag {\em does} give proton helicity flip amplitudes at leading
twist, with an overall factor of $\dt /m$, or $\dt R_p$ where $R_p$ is
the proton radius.  While with the constraints from parity invariance
there are in general 12 independent helicity amplitudes in VCS
\cite{gui} the handbag model leaves us with four: the combination of
the helicities of the photon, incoming and outgoing proton gives $2
\times 2 \times 2$ amplitudes, which are related pairwise by a parity
transformation. These four independent amplitudes can be expressed in
terms of the four off-diagonal parton distributions introduced
in~\cite{Ji}.

We have so far only considered the Born diagrams in the handbag
approach, and now have to discuss the effects of QCD loop corrections.
For diagrams with quark legs attached to the soft blob our above
helicity arguments are not changed, as perturbative corrections to the
hard scattering respect quark helicity conservation. For the diagrams
that involve off-diagonal gluon correlations we have to consider the
collinear scattering of a photon on a free gluon. As on-shell gluons
are transverse one finds that this scattering can change the photon
helicity by zero or two units, but not by one. The virtual photon can
thus not be longitudinal since the final state photon must be
transverse, but contrary to the quark scattering case there can now be
photon helicity transitions from $- 1$ to $+1$ and vice versa. Photon
helicity conservation can thus be violated at the level of
$\alpha_S$-corrections. We will keep this in mind when discussing
tests of the handbag in the following. The measurement of such $- 1$
to $+1$ helicity transitions would in fact be interesting by itself,
since they involve gluon correlation functions that have no
counterpart in the diagonal limit $p$ = $p'$, where they are forbidden
by angular momentum conservation.

\begin{figure}
  $$\epsfysize 4cm\epsfbox{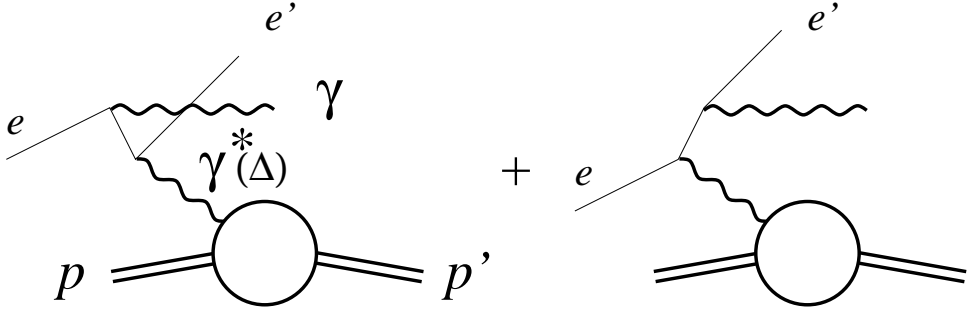}$$
  \caption{\label{BetheHeitler}The Bethe-Heitler process.}
\end{figure}

\vspace{\baselineskip}
\noindent
4. In the electroproduction process
\begin{equation}  \label{ep}
  e(k) + p(p) \to e(k') + p(p') + \gamma(q') \eqcm
\end{equation}
virtual Compton scattering interferes with the Bethe-Heitler (BH)
process of Fig.~\ref{BetheHeitler}. We use the
variables $Q^2$, $\xbj$ and $\dt$ (or $t$) already introduced and
further $y = (q \cdot p) / (k \cdot p)$. Another convenient variable
is the ratio $\epsilon$ of longitudinal to transverse initial photons
in the VCS process. For reasons that will be clear shortly we assume
in the following that
\begin{equation}
  \label{restricty}
  1 - y \gg m^2 /Q^2  \eqcm
\end{equation}
then $\epsilon = (1 - y) /(1 - y + y^2 /2)$.

\begin{figure}
  $$\epsfysize 6cm\epsfbox{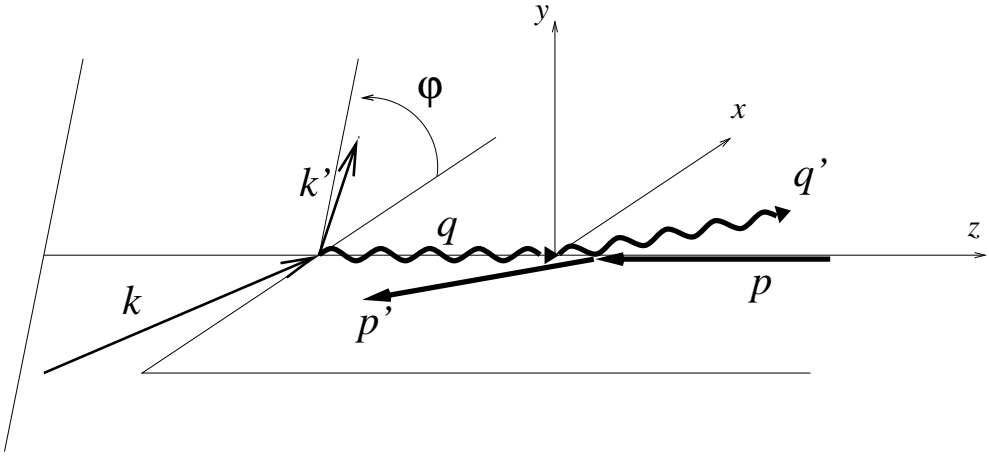}$$
  \caption{\label{kinematics}Kinematics of reaction \protect(\ref{ep})
    in the c.m.\ of the scattered photon and proton.}
\end{figure}

We also introduce the angle $\varphi$ between the leptonic and
hadronic scattering planes in the c.m.\ of the scattered photon and
scattered proton (see Fig.~\ref{kinematics}). As we will show, the
$\varphi$-dependence of the physical cross section contains a wealth
of information. Rather than work at fixed $\varphi$ values --- a
procedure sometimes advocated to knock down the Bethe-Heitler
contribution --- we will show how to test the model using the full
$\varphi$-dependence. Notice that with appropriate phase conventions
for the external particles the contribution of VCS to the amplitude
of~(\ref{ep}) for a given helicity $\lambda$ of the intermediate
photon $\gstar(q)$ has the $\varphi$-dependence
\begin{equation}
  \label{VCSamplitude}
  \tvc \propto \exp(- i \lambda \varphi)  \eqcm
\end{equation}
which will allow us to find simple tests for the helicity structure of
VCS.

Let us take a closer look at the BH process in the kinematic region
where we want to test the handbag approximation, i.e.\ where $\dt$,
$\sqrt{-t}$ and $m$ are all small compared with $Q$. Since the handbag
dominance holds in a $1/Q$ expansion, it is natural to also expand the
BH amplitude in powers of $1/Q$. Its leading term goes like $\dt/t$ or
$m/t$, whereas $\tvc$ scales like $1 /Q$ if the $\gstar p \to \gamma
p$ amplitude behaves in $Q$ like a constant.\footnote{We assume that
  none of the $\gstar p\to\gamma p$ helicity amplitudes {\em
    increases} like a power of $Q$.} This leads to a hierarchy in
powers of $1 /Q$ for the contributions to the $ep$ cross section of
the squared BH amplitude, the VCS--BH interference and the squared VCS
amplitude, respectively.

To understand the $\varphi$-dependence of the BH amplitude it is worth
noting that the invariants $s' = (k' + q')^2$ and $u' = (k - q')^2$,
which appear in the lepton propagators at Born level, read
\begin{eqnarray}
\label{sbh}
s' &=&   {Q^2 \over y}
       - {2 \dt Q \sqrt{1-y} \over y} \cos \varphi
       + O(m^2, t)  \eqcm \\
\label{ubh}
u' &=& - {Q^2 \over y} (1-y)
       + {2 \dt Q \sqrt{1-y} \over y} \cos \varphi
       + O(m^2, t)  \eqpt
\end{eqnarray}
We see that $s'$ and $u'$ depend only weakly on $\varphi$ in the
region of interest. Under the condition (\ref{restricty}) we can
expand in powers of $(\dt /Q) \cdot \cos\varphi$ the propagators $1 /
s'$ and $1 / u'$ in the expression of $\tbh$, which gives at the same
time a Fourier expansion in $\varphi$. To leading order in $1 /Q$ we
find a very simple $\varphi$-dependence:
\begin{equation}
  \label{BHamplitude}
  \tbh \propto \exp(- 2 i \lambda' \varphi)  + O\left( 1 \over Q
  \right)
\end{equation}
for a scattered photon of helicity $\lambda'$.

We can now investigate the $\varphi$-dependence of the $ep$ cross
section, with unpolarized protons and leptons for the time being. It
is given by
\begin{equation}
  \label{crosssection}
{d\sigma \over d\varphi \, d t \, d Q^2 \, d\xbj} =
  \frac{1}{32 \, (2 \pi)^4} \, \frac{\xbj \, y^2}{Q^4} \, \frac{1}{
  \sqrt{1 + 4 \xbj^2 m^2 /Q^2}} \,
  | \tbh + \tvc |^2  \eqpt
\end{equation}
The squared BH contribution has the structure
\begin{equation}
  \label{BHsquared}
  | \tbh |^2 = f_1(\epsilon, \xbj, \dt) + \frac{1}{Q} \cos\varphi
  \cdot f_2(\epsilon, \xbj, \dt) + O\left( 1 \over Q^2 \right)
  \eqcm
\end{equation}
with functions $f_1$ and $f_2$ whose expressions we do not need here.
The square of VCS can readily be expressed in terms of helicity
amplitudes $\M{\lambda}{\lambda'}{h}{h'}(Q^2, \xbj, \dt)$ for $\gstar
p \to \gamma p$, where $\lambda$ ($\lambda'$) is the helicity of the
initial (final) state photon and $h$ ($h'$) that of the initial
(final) state proton. It reads
\begin{eqnarray}
  \label{VCSsquared}
| \tvc |^2 &=& \frac{e^6}{Q^2} \,
\frac{2}{1- \epsilon} \, \sum_{h, h'}  \left[ \:
  {1 \over 2} \{ | \M{1}{1}{h}{h'} |^2 +
                 | \M{- 1}{1}{h}{h'} |^2 \}
   + \epsilon \, | \M{0}{1}{h}{h'} |^2 \right. \nonumber \\
&& {} -  \cos\varphi \cdot \sqrt{\epsilon (1 + \epsilon)} \, \re
    \{ (\M{1}{1}{h}{h'})^{*} \M{0}{1}{h}{h'}
     - (\M{- 1}{1}{h}{h'})^{*} \M{0}{1}{h}{h'}
    \} \nonumber \\
&& \left. {} - \cos 2 \varphi \cdot \epsilon \, \re
    \{ (\M{1}{1}{h}{h'})^{*} \M{- 1}{1}{h}{h'}
    \} \rule{0em}{3ex}  \: \right]  \eqcm
\end{eqnarray}
where $e$ is the electron charge. Finally, thanks to the simple
$\varphi$-behavior in~(\ref{VCSamplitude}),~(\ref{BHamplitude}) the
BH--VCS interference can be written as
\begin{eqnarray}
  \label{interference}
\tbh^{*} \tvc^{} + \tvc^{*} \tbh^{} &=&
  \frac{e^6}{t} \, \frac{m}{Q} \cdot \frac{4 \sqrt{2}}
  {\xbj\sqrt{1 - \xbj}}
\left[ \cos\varphi \cdot \frac{1}{\sqrt{\epsilon (1 - \epsilon)}} \,
           \re \MM{1}{1} \right. \nonumber \\
&& \hspace{-12em} \left. {} - \cos 2\varphi \cdot
   \sqrt{ \frac{1 + \epsilon}{1 - \epsilon} } \, \re \MM{0}{1}
             - \cos 3\varphi \cdot
   \sqrt{ \frac{\epsilon}{1 - \epsilon} } \, \re \MM{-1}{1} \right]
 + O\left( 1 \over Q^2 \right) \eqcm
\end{eqnarray}
for a negatively charged lepton. Here the $\gstar p \to \gamma p$
amplitudes enter in linear combinations
\begin{eqnarray}
  \label{combinations}
\MM{\lambda}{\lambda'}(Q^2, \xbj, \dt) &=&
   \frac{\dt}{m} \cdot \Big[ (1 - \xbj) \, G_M
                 - (1 - \xbj/2) \, F_2 \Big] \cdot
  \M{\lambda}{\lambda'}{- 1/2}{- 1/2} \nonumber \\
& + & \frac{\dt}{m} \cdot \Big[ G_M - (1 - \xbj/2) \, F_2 \Big] \cdot
  \M{\lambda}{\lambda'}{1/2}{1/2} \nonumber \\
& + & \left[ \xbj^2 \, G_M + \frac{\dt^2}{2 m^2} \, F_2 \right] \cdot
  \M{\lambda}{\lambda'}{1/2}{- 1/2} \nonumber \\
& - & \frac{\dt^2}{2 m^2} \, F_2 \cdot
  \M{\lambda}{\lambda'}{- 1/2}{1/2}  \eqcm
\end{eqnarray}
where $G_M$ is the magnetic and $F_2$ the Pauli form factor of the
proton, both to be taken at momentum transfer $t$.

In summary, then, the hierarchy in powers of $1/Q$, with $| \tbh |^2$,
BH--VCS interference and $| \tvc |^2$ going like 1, $1/Q$ and $1/Q^2$,
respectively, in the handbag model, is accompanied with a
$\varphi$-distribution going like $\cos (n \varphi)$ with $n = 0$, 1,
2 and 3. A powerful consequence is that extracting the angular
distribution does not require any binning of data. One can instead
weight the events with these angular functions, i.e.\ measure
weighted cross sections
\begin{equation}
  \label{weighted}
\weight{f(\varphi)} = \frac{Q^4}{\xbj \, y^2} \, \int d\varphi \,
  {d\sigma \over d\varphi \, d t \, d Q^2 \, d\xbj} \cdot f(\varphi)
\end{equation}
with $f(\varphi)= 1, \cos\varphi, \cos 2\varphi, \cos 3\varphi$, which
directly project out the corresponding Fourier coefficients in the
cross section. In $\weight{f(\varphi)}$ we have taken out the factor
$\xbj \, y^2 / Q^4$ from the phase space in (\ref{crosssection}) for
later convenience. Other choices of weighting functions are possible:
one can e.g.\ multiply the data by $\pm 1$ in alternating sectors,
generalizing the concept of a left-right asymmetry, which is
tantamount to choosing $f(\varphi)$ as the signature function of
$\cos\varphi$, $\cos 2\varphi$, \ldots\ Due to constraints from
angular acceptance in a given experiment one may also have to choose
functions $f(\varphi)$ that vanish in a certain range of $\varphi$. In
any case one can use the statistics of the full data set to extract
the Fourier coefficients of the cross section.

In (\ref{VCSsquared}) and (\ref{interference}) we note the very
different $\epsilon$-dependence of the VCS and BH amplitudes: $\tvc$
has a relative factor of $1 / \sqrt{1 - \epsilon}$ compared with
$\tbh$. The numerical importance of $\tvc$ might thus be enhanced by
exploiting the region close to $\epsilon = 1$. At fixed $Q^2$ and
$\xbj$ this requires a large c.m.\ energy of the $ep$ reaction.
Conversely, we note in the interference term (\ref{interference}) that
small $\epsilon$ emphasizes the first term in the square brackets,
going like $\M{1}{1}{h}{h'}$, which is one of the amplitudes we want
to test.

Using the weighted cross sections (\ref{weighted}) we now formulate a
number of tests for the predictions of the handbag approximation. Let
us first assume that $\epsilon$ is not too large so that the BH--VCS
interference (\ref{interference}) dominates over the squared VCS
amplitude (\ref{VCSsquared}) due to their respective global factors $1
/Q$ and $1 /Q^2$.
\begin{itemize} 
\item We can test the scaling properties of the handbag: in the
  leading order interference term~(\ref{interference}) the function
  $\cos\varphi$ is multiplied by the photon helicity conserving
  amplitudes $\M{1}{1}{h}{h'}$ which the handbag approximation
  predicts to be constant in $Q$. There is also a $\cos\varphi$-term
  in the square~(\ref{BHsquared}) of the BH amplitude, which has the
  same global power $1 /Q$ and thus must be subtracted if we want to
  investigate VCS. Note that the BH process including its QED
  radiative corrections can be calculated and that the elastic proton
  form factors are well parameterized in the region of small $t$ where
  they are needed. We then have as a test for the scaling properties
  in the handbag that at fixed $\dt$, $\xbj$ and $y$
  \begin{equation}
    \label{cos}
    \weight{\cos\varphi} - 
    \left.\weight{\cos\varphi}\right._{\rm BH \ only} \sim 1 /Q  \eqcm
  \end{equation}
  where $\smash{ \left.\weight{\cos\varphi}\right._{\rm BH \ only} }$
  denotes the contribution of the squared BH amplitude. Of course this
  scaling behavior is to be understood as up to logarithms due to QCD
  radiative corrections in the handbag.
  
  Note that with a fixed $ep$ c.m.\ energy the variables $\xbj$, $y$
  and $Q^2$ are not independent: changing $Q^2$ at fixed $\xbj$ will
  change $y$. To investigate the $Q^2$-behavior of the amplitudes
  $\M{1}{1}{h}{h'}$ one must then multiply $\weight{\cos\varphi}$ with
  an appropriate $\epsilon$-dependent factor to be taken from
  (\ref{interference}). A similar remark holds for the other weighted
  cross sections we will discuss.
  
  Beyond testing the handbag the measurement of $\weight{\cos\varphi}$
  should offer a good way to extract the leading twist VCS amplitudes,
  in a linear combination~(\ref{combinations}). This allows one to
  gain rather direct information on the new off-diagonal parton
  distributions if the handbag approximation is found to work well in
  a given kinematic regime.
\item To test photon helicity conservation we can use $\weight{\cos
    2\varphi}$ and $\weight{\cos 3\varphi}$. To order $1 /Q$ they
  receive contributions from the BH--VCS interference proportional to
  $\M{0}{1}{h}{h'}$ and $\M{-1}{1}{h}{h'}$, respectively, which are
  zero in the handbag approximation and thus should be power
  suppressed. Note that $| \tbh |^2$ does not contain any $\cos
  2\varphi$ or $\cos 3\varphi$ up to order $1 /Q$, so that, {\sl
    without needing to subtract this contribution}, we have as a test
  for the handbag that
  \begin{equation}
    \label{costwothree}
    \weight{\cos 2\varphi} \eqcm \weight{\cos 3\varphi} \sim 1 / Q^n
    \eqcm n \ge 2  \eqpt
  \end{equation}
  From (\ref{cos}) and (\ref{costwothree}) one has of course that
  $\weight{\cos 2\varphi}$ and $\weight{\cos 3\varphi}$ are small
  compared with $\weight{\cos\varphi}$, a test that can even be done
  without much lever arm in $Q^2$.
  
  At the end of Sec.~3 we discussed the possibility of having leading
  twist amplitudes $\M{-1}{1}{h}{h'}$ at order $\alpha_S$ in the
  handbag. If such amplitudes exist then the behavior of
  $\weight{\cos 3\varphi}$ should be like $\alpha_S / Q$
  instead of the one given in (\ref{costwothree}).
\end{itemize}
Let us now see what can be done in the kinematic regime where
$\epsilon$ is sufficiently close to 1 so that the VCS contribution to
the amplitude dominates over the contribution from BH. It is then $|
\tvc |^2$ that is dominant in the cross section.
\begin{itemize}
\item The scaling of the handbag can now be tested with the weighted
  cross section $\weight{1}$: the constant part in the squared VCS
  amplitude (\ref{VCSsquared}) contains a term quadratic in the photon
  helicity conserving amplitudes $\M{1}{1}{h}{h'}$.  We remark that
  the leading order interference term given in~(\ref{interference})
  does not contribute to $\weight{1}$, but that the terms denoted by
  $O(1 /Q^2)$ there do contain a constant piece. By assumption it is
  suppressed compared with the contribution from $| \tvc |^2$, as well
  as the leading term in the squared BH amplitude~(\ref{BHsquared}),
  which one may but does not need to subtract. The scaling prediction
  of the handbag model is then
  \begin{equation}
    \label{const}
    \weight{1} \sim 1 /Q^2    \eqpt
  \end{equation}
\item $\weight{\cos 2\varphi}$ is sensitive to the photon helicity
  changing amplitudes $\M{- 1}{1}{h}{h'}$. The quark contribution 
  in the handbag approximation gives  
  \begin{equation}
    \label{costwo}
    \weight{\cos 2\varphi} \sim 1 / Q^n    \eqcm n \ge 3 \eqcm
  \end{equation}
  whereas the gluon diagrams could enhance this behavior by a factor 
  $\alpha_S \cdot Q$. There is a corresponding
  test for the amplitudes $\M{0}{1}{h}{h'}$ involving
  $\weight{\cos\varphi}$, but the contribution of the BH--VCS
  interference to this quantity is proportional to the leading twist
  amplitudes $\M{1}{1}{h}{h'}$, so that one will need smaller values
  of $(1 - \epsilon)$ to suppress this contribution than in the case of
  $\weight{\cos 2\varphi}$.
\end{itemize}

\vspace{\baselineskip}
\noindent
5. Let us now see which additional information on VCS can be gained
using lepton beams with longitudinal polarization $h_e = \pm 1/2$,
while still averaging over the proton spin. Recall that due to parity
and time reversal invariance the $h_e$-dependent part of the cross
section is the interference between the absorptive and the
nonabsorptive part of the $ep$ scattering
amplitude~\cite{KSG}.\footnote{Remember that, apart from kinematical
  phases such as in~(\ref{VCSamplitude}),~(\ref{BHamplitude}) and from
  phases due to the definition of particle states the absorptive
  (nonabsorptive) part of an amplitude is its imaginary (real) part.}
The BH amplitude is purely nonabsorptive at Born level. QED radiative
corrections, although they can be sizeable, are not expected to
significantly change this since a large part of them comes from the
collinear and infrared regions which do not lead to large absorptive
parts. As a consequence $| \tbh |^2$ is independent of $h_e$. The
$h_e$-dependent terms in the BH--VCS interference probe the absorptive
part of VCS, i.e.\ $\im \M{\lambda}{\lambda'}{h}{h'}$ with our phase
conventions, while its $h_e$-independent part is sensitive only to
$\re \M{\lambda}{\lambda'}{h}{h'}$.

From parity invariance it follows that for unpolarized protons the
$h_e$-dependent part of the cross section is odd in $\varphi$ and the
$h_e$-independent part is even. The lepton spin asymmetry of the
$\varphi$-integrated cross section is therefore zero, which might
offer a useful experimental cross check for the lepton spin
measurement and the acceptance in $\varphi$.

Compared with the lepton spin averaged expressions
in~(\ref{VCSsquared}),~(\ref{interference}), we now have
\begin{eqnarray}
  \label{VCSsquaredspin}
  | \tvc |^2 &=&
  \left\{ | \tvc |^2 \right\}_{\rm eq.(\ref{VCSsquared})} +
 \frac{e^6}{Q^2} \, \frac{2}{1- \epsilon} \cdot  \nonumber \\
&& \sum_{h, h'} 2 h_e \, \sin\varphi \cdot \sqrt{\epsilon (1 -
  \epsilon)} \, \im
    \{ (\M{1}{1}{h}{h'})^{*} \M{0}{1}{h}{h'}
     - (\M{- 1}{1}{h}{h'})^{*} \M{0}{1}{h}{h'}
    \}
\end{eqnarray}
and
\begin{eqnarray}
  \label{interferencespin}
  \tbh^{*} \tvc^{} + \tvc^{*} \tbh^{} &=&
  \left\{ \tbh^{*} \tvc^{} + \tvc^{*} \tbh^{} \right\}_{\rm
      eq.(\ref{interference})} +
  \frac{e^6}{t} \, \frac{m}{Q} \cdot \frac{4 \sqrt{2}}
  {\xbj\sqrt{1 - \xbj}} \cdot \nonumber \\
&& \hspace{-10em}  2 h_e \left[
  - \sin\varphi \cdot \sqrt{ \frac{1 + \epsilon}{\epsilon} } \,
  \im \MM{1}{1} + \sin 2\varphi \cdot \im \MM{0}{1} \right]
 + O\left( 1 \over Q^2 \right)  \eqpt
\end{eqnarray}
Note the absence of $\sin 3\varphi$ in the interference term and of
$\sin 2\varphi$ in $| \tvc |^2$. We can now complete our list of tests
for the handbag by considering two more weighted cross sections,
starting again with the region of $\epsilon$ where the BH--VCS
interference dominates over the square of VCS :
\begin{itemize}
\item $\weight{\sin\varphi}$ is analogous to $\weight{\cos\varphi}$
  discussed above, with the difference that now no subtraction of $|
  \tbh |^2$ is necessary. The scaling prediction of the handbag is
  \begin{equation}
    \label{sin}
    \weight{\sin\varphi} \sim 1 /Q  \eqpt
  \end{equation}
  We emphasize again that $\weight{\sin\varphi}$ and $\smash{
    \left.\weight{\cos\varphi}-\weight{\cos\varphi}\right._{\rm BH\ 
      only}}$ offer a way to extract the imaginary and real parts of
  the $\M{\lambda}{\lambda'}{h}{h'}$ separately. This is a great
  advantage of the interference term compared with $| \tvc |^2$, where
  real and imaginary parts mix and should be difficult to disentangle.
  Remember that owing to P and T invariance the off-diagonal parton
  distributions introduced in \cite{Ji} are real valued, so that
  $\smash{ \im \M{\lambda}{\lambda'}{h}{h'} }$ is given by the
  off-diagonal distributions at $x = \xbj$ and $x' = 0$ and $\smash{
    \re \M{\lambda}{\lambda'}{h}{h'} }$ by a principal value integral
  over $x$. Separating the real and imaginary parts of $\smash{
    \M{\lambda}{\lambda'}{h}{h'} }$ is crucial to obtain detailed
  information on the new parton distributions.
\item $\weight{\sin 2\varphi}$ only receives contributions from the
  interference. As $\weight{\cos 3\varphi}$ and $\weight{\cos
    2\varphi}$ it can be used to test for photon helicity
  conservation, the handbag prediction being
  \begin{equation}
    \label{sintwo}
    \weight{\sin 2\varphi} \sim 1 / Q^n \eqcm n \ge 2  \eqpt
  \end{equation}
  We should remark that these quantities test for different photon
  helicity changing amplitudes and are thus not redundant:
  $\weight{\sin 2\varphi}$ is sensitive to $\im \M{0}{1}{h}{h'}$,
  $\weight{\cos 2\varphi}$ to $\re \M{0}{1}{h}{h'}$, and
  $\weight{\cos 3\varphi}$ to $\re \M{- 1}{1}{h}{h'}$.
\end{itemize}
For large $\epsilon$ where the square of VCS dominates the cross
section one might use $\weight{\sin\varphi}$ to test for the
amplitudes $\M{0}{1}{h}{h'}$, with the same caveat concerning the
BH--VCS interference contribution we have made for
$\weight{\cos\varphi}$ in the previous section.

We finally remark that the measurement of $\weight{\sin\varphi}$ and
$\weight{\sin 2\varphi}$ does {\em not} require us to form a lepton
spin asymmetry. In fact, all moments~(\ref{cos}) --~(\ref{costwo})
and~(\ref{sin}) --~(\ref{sintwo}) can be determined independently with
just one nonzero value of the lepton polarization.

\vspace{\baselineskip}
\noindent
6. We have shown how the {\em photon} spin structure of $\gstar p \to
\gamma p$ can be investigated using the distribution of the angle
$\varphi$, both with and without lepton polarization. To measure the
dependence of $\M{\lambda}{\lambda'}{h}{h'}$ on the {\em proton}
helicities $h$, $h'$ one needs an experiment with proton
polarization. A different class of tests of the handbag dominance can
be built along the lines drawn in the present study.

We shall not elaborate on this subject here, but just illustrate
what are the additional possibilities brought by the extra degree
of freedom, focusing on the amplitudes where the proton changes
its helicity. A possibility to extract information on
these amplitudes is offered by a transversely polarized proton
target, the transverse spin asymmetry being given by the
interference of proton helicity flip with helicity no-flip amplitudes.
In the handbag model the proton helicity flip amplitudes have a factor
$\dt R_p$ as mentioned in Sec.~3. That is, they are linear in $\dt$
for $\dt \to 0$ as required by angular momentum conservation for
amplitudes where the photon helicity does not change. In general there
is however an amplitude $\M{0}{1}{1/2}{- 1/2}$, which can be finite at
$\dt = 0$ since both proton and photon change spin by one unit. This
amplitude is zero in the handbag because the photon is
longitudinal. We then have as a further prediction that to leading
order in $1 /Q$ proton helicity flip amplitudes must vanish like $\dt$
as $\dt$ goes to zero.

\vspace{\baselineskip}
\noindent
{\bf Acknowledgments}. \\
We gratefully acknowledge discussions with Mauro Anselmino, John
Collins, Pierre Guichon, Peter Kroll, Peter Landshoff, Wei Lu and Otto
Nachtmann. T. G. was carrying out his work as part of a training
Project of the European Community under Contract No. ERBFMBICT950411.
This work has been partially funded through the European TMR Contract
No.~FMRX-CT96-0008: Hadronic Physics with High Energy Electromagnetic
Probes, and through DOE grant No.~85ER40214.

\end{document}